\pdfoutput=1
\documentclass[numbers]{article}
\pdfoutput=1
\usepackage[english]{babel}
\usepackage[letterpaper,top=2cm,bottom=2cm,left=3cm,right=3cm,marginparwidth=1.75cm]{geometry}
\usepackage[colorlinks=true,urlcolor=blue,anchorcolor=blue,citecolor=blue,filecolor=blue,linkcolor=blue,menucolor=blue,pagecolor=blue,linktocpage=true,pdfproducer=medialab,pdfa=true]{hyperref}
\usepackage{jcappub}
\PassOptionsToPackage{sort&compress,round,numbers}{natbib}
\usepackage{lmodern}
\usepackage{float}
\usepackage{amsfonts}
\usepackage{subcaption}
\usepackage{esint}
\usepackage{xcolor}
\usepackage{cancel}
\usepackage{soul}
\hypersetup{colorlinks = true, urlcolor = blue, linkcolor = blue, frenchlinks = true, pdfborder = {0 0 0}}
\setstcolor{red}
\setcitestyle{square}
\allowdisplaybreaks
\usepackage{amsmath}
\usepackage{amsthm}
\usepackage{amssymb}
\usepackage{graphicx}
\newtheorem{lemma}{Lemma}
\title{Synge's world function applied to causal diamonds and causal sets}
\author{Arad Nasiri}
\affiliation{Theoretical Physics Group, Blackett Laboratory, Imperial College London, SW7 2AZ, UK}
\emailAdd{a.nasiri21@imperial.ac.uk}

\begin{document}

\abstract{One of the major tasks in discrete theories of gravity, including causal set theory, is to discover how the combinatorics of the underlying discrete structure recovers various geometric aspects of the emergent spacetime manifold. In this paper, I develop a new covariant approach to connect the combinatorics of a Poisson sprinkled causal set to the geometry of spacetime, using the so-called Synge's world function. The Poisson sprinkling depends crucially on the volume of a causal interval. I expand this volume, in 2 dimensions, in powers of Synge's world function and Ricci scalar. Other geometric properties of Synge's world function are well-known, making it easy to connect to the curvature of spacetime. I use this connection to provide a straightforward proof that the BDG action of causal sets gives the Einstein-Hilbert action in the continuum limit, without having to work in a particular coordinate system like Riemann normal coordinates.}
\maketitle
\section{Introduction}

Causal set theory (CST) is an approach to quantum gravity that takes causality, discreteness, and objectivity of spacetime events as its fundamental premises~\cite{bombelli1987space,surya2019causal,sorkin2005causal,surya2011directions}. Starting with a continuous manifold of spacetime, we can sample a set of discrete spacetime point-events with a Poisson distribution. The fixed density of this random sampling is supposedly close to the Planck density. We then throw the causal structure of spacetime ($\preceq$, for which $x\preceq y$ indicates the existence of a causal curve from $x$ to $y$) onto the sprinkled points, and then we forget about the initial manifold. We change our viewpoint, and treat the countable partially ordered set -a causal set- as fundamental. Also, note that the random nature of such a sampling is of great importance. It guarantees that the sprinkling will not break Lorentz invariance \cite{bombelli2009discreteness,dowker2020symmetry}. 
In a Poisson sprinkling, the probability of having $n$ elements in a region of spacetime with volume $V$ is 
\begin{equation}
    P(n)=\frac{1}{n!}e^{-\rho V}(\rho V)^n
\end{equation}
where $\rho$ is the sprinkling density. The expected number of points in this region is $\langle n\rangle=\rho V$; this is the number-volume correspondence in causal sets. It turns out that under some assumptions, the Poisson distribution has the smallest variance from this correspondence among all random point processes \cite{saravani2014causal}.\\

Now imagine a particular causal set $(\mathcal{C},\prec)$ is given. If the fundamental fabric of spacetime is such a discrete structure, the spacetime as a continuous manifold would only be an emergent structure: The idea is that at scales much larger than the discreteness scale, the causal set can be approximated by a continuous manifold. But what does this approximation, this correspondence, mean? One way to think about it is to say that a Lorentzian manifold $(\mathcal{M},g)$ emerges out of the given causal set if $\mathcal{C}$ is a “typical" outcome of a Poisson sprinkling on $\mathcal{M}$.

Now a natural question concerns the uniqueness of this correspondence. Of course, there are many “typical" Poisson sprinkled causal sets associated with a given manifold, just as there are many “typical" random bit strings. However, the uniqueness from the other side is important: Is the manifold $\mathcal{M}$ unique in some sense? If it is not unique at all, it would be a serious problem. After all, by replacing a smooth manifold by a discrete causal structure, our idea was to represent the physical content of GR using a more fundamental graph-like structure. However, if two manifolds $\mathcal{M}_1$ and $\mathcal{M}_2$ that are geometrically distinct on scales larger than the discreteness scale can both approximate a given causal set $\mathcal{C}$ in the above sense, then it means that the partial order of a causal set does not have enough data to represent the geometry of a Lorentzian manifold; at least two different large-scale geometric predictions can be made from $\mathcal{C}$. It would be like having an atomic view of matter that can give rise to two distinct sets of thermodynamical parameters for the same atomic state. In such a situation, a statistical theory would not be thought of as the basis for thermodynamics.

A first hint that a causal set is indeed enough to recover a unique Lorentzian manifold is provided by the so-called HKMM theorem \cite{surya2019causal,hawking1976new,malament1977class}. Roughly, it states that if two Lorentzian manifolds have the same causal structure, in the sense of an order-preserving bijection $f$ existing between them, then the topology and differentiable structure of the two manifolds are the same, and they are isometric up to a conformal factor. Indirectly, this is pointing toward a uniqueness theorem. The reason is that if both $\mathcal{M}_1$ and $\mathcal{M}_2$ have $\mathcal{C}$ as a typical causal set from Poisson sprinkling, then there is a mapping $f$ from the embedding of $\mathcal{C}$ inside $\mathcal{M}_1$ to its embedding in $\mathcal{M}_2$, that preserves the causal structure of the two manifolds induced on the embedded points. Then, roughly speaking, the causal structure of the two spacetimes must be the same down to the discreteness scale. In addition, from the number-volume correspondence, the volume measures of the two manifolds are so close that for regions $U_1\subset \mathcal{M}_1$ and $U_2\subset \mathcal{M}_2$ that contain exactly the same embedded points of $\mathcal{C}$, $$\frac{\text{vol}_{\mathcal{M}_1}(U_1)-\text{vol}_{\mathcal{M}_2}(U_2)}{\text{vol}_{\mathcal{M}_1}(U_1)}\sim \frac{1}{\sqrt{\rho\ \text{vol}_{\mathcal{M}_1}(U_1)}}.$$
Under these circumstances, we expect a theorem, analogous to HKMM, that enforces $\mathcal{M}_1$ and $\mathcal{M}_2$ to be topologically equivalent and isometric at scales large compared to the discreteness scale. This is the fundamental conjecture of causal sets: the \textit{Hauptvermutung}. The precise formulation of an “isometry at large scales” is yet to be carried out. Nevertheless, the evidence for the correctness of the conjecture has been adding up over years.

In this regard, one of the main directions of research in CST has been to recover the geometric properties of the spacetime from the pure combinatorics of the underlying causal set. For recovering the metric, or in fact the timelike and spatial distances, see \cite{brightwell1991structure,rideout2009spacelike}. Of great importance is the recovery of curvature. The full Riemann tensor has not been worked out yet; however, the combinatorial version of the Ricci scalar has been found. This has a two-fold significance: Not only does it say that a partial order can encode geometrical information, but also it provides the analogue of the Einstein-Hilbert action for the causal sets. This is the Benincasa-Dowker-Glaser (BDG) action \cite{benincasa2010scalar,belenchia2016continuum,benincasa2013action}. Its derivation was inspired by a totally different line of thought in CST; namely the dynamics of a scalar field on a causal set.\\

As an attempt to investigate the evolution of a massless scalar field on a causal set, instead of starting from an action principle for the scalar field, Sorkin started directly from the equation of motion \cite{sorkin2009does}. In other words, he tried to find a causal set analogue for the d'Alembertian operator. Since a priori, there is no such thing as a tangent space or differential operator as a vector field on a causal set, linear operators like $\Box$ acting on a field $\phi$ can only be thought of as being a linear combination of the values of $\phi$ at different causal set elements. In the same spirit, for a function $f$ defined on a grid over the real line, the closest thing to $f'(x_i)$ is the linear combination $\left(f(x_i+l)-f(x_i)\right)/l$, $l$ being the grid spacing. So similarly, for a 2d causal set $\mathcal{C}$, Sorkin defines the retarded operator
\begin{equation}
\label{Bphi0}
    B\phi(z):=\frac{4}{l^2}\left(-A\phi(z)+B\sum_{y:|[y,z]|=2}\phi(y)-C\sum_{y:|[y,z]|=3}\phi(y)+D\sum_{y:|[y,z]|=4}\phi(y)\right),
\end{equation}
where $[y,z]=\{x\in\mathcal{C}|y\preceq x\preceq z\}$ is the inclusive order interval between $y$ and $z$. $l$ is the discreteness length.
The first thing to note is that to ensure causality is preserved, the linear combination only runs over those elements that are in the causal past of $z$. Secondly, this is a layered sum: All the field values at elements that have the same number of elements between them and $z$, appear with the same coefficient. One could in principle continue this summation for deeper layers in the past, but it turns out that for 2 dimensions, 3 layers are enough to get the d'Alembertian operator. To connect this discrete version of the d'Alembertian operator to the usual continuous version, Sorkin requires that for causal sets that are Poisson sprinkled in a flat 2-dimensional manifold with density $\rho=1/l^2$
\begin{equation}
\label{EB}
    \lim_{\rho\rightarrow\infty}\mathbb{E}[B\phi](z)=\Box \phi(z),
\end{equation}
where $\phi$ is a scalar field of compact support on the 2d Minkowski spacetime, the expectation value is with respect to the Poisson sprinkling process on the manifold, and $z$ is added to each sprinkled causal set by hand. \eqref{EB} fixes the coefficients \begin{equation}
    A=\frac{1}{2},\ \ B=1, \ \ C=2,\ \ D=1.
\end{equation}Sorkin proves this by writing a Taylor expansion for the field $\phi$ around $z$ \cite{skpr}. I shall give a new approach for computing the above expectation value and finding the coefficients in section \ref{flatflt}. It will have the benefit of revealing an exact integral equation relating $\Box\phi$ and $\mathbb{E}[B\phi]$ at finite density.\\

Now what about curved spacetime? It turns out that the Ricci scalar enters the expectation value of the causal set box operator, and that gives the motivation to use $B$ to define the analogue of Einstein-Hilbert action for causal sets. More specifically, it turns out that in curved spacetime,
\begin{equation}
\label{b-r}
    \lim_{\rho\rightarrow\infty}\mathbb{E}[B\phi](z)=(\Box-\frac{R}{2})\phi(z).
\end{equation}
Therefore, if one chooses to set $\phi=-1$, then this motivates the causal set “Lagrangian" in 2d:
\begin{equation}
    \mathcal{L}(z)=2\rho \left(1\ -\ 2\sum_{y:|[y,z]|=2}1\ +\ 4\sum_{y:|[y,z]|=3}1\ -\ 2\sum_{y:|[y,z]|=4}1\right),
\end{equation}
since $\lim_{\rho\rightarrow\infty}\mathcal{L}(z)=R(z)/2$. Then summing $\mathcal{L}(z)$ over all causal set points gives us the BDG action for any causal set $\mathcal{C}$:
\begin{equation}
    \frac{S_{BDG}}{\hbar}=2\left(N-2N_1+4N_2-2N_3\right),
\end{equation}
where,
\begin{equation}
    N_i=\left|\ \{(x,y):\ x,y\in\mathcal{C};\ x\preceq y,\ |[x,y]|=i+1\}\ \right|,
\end{equation}
\begin{equation}
    N=|\mathcal{C}|.
\end{equation}

I aim to prove (\ref{b-r}) (for constant $\phi$) with the technology of Synge's world function. The existing proof in the literature \cite{belenchia2016continuum} relies on writing everything down in Riemann normal coordinates. In this paper, I show that this is also possible in a fully covariant manner, without having to resort to any particular coordinate system. The spirit of the computation is similar to what is done for flat spacetime in section \ref{flatflt}, but one uses the geometric properties of Synge's world function and the volume of causal intervals for curved manifolds. Therefore, I review the basics of Synge's world function in section \ref{Syngesin}. Next I expand the volume of a causal interval (also known as a causal diamond or Alexandrov interval) in terms of the powers of Synge's world function and $R$ in section \ref{volvo}. The two methods used in the literature  \cite{myrheim1978statistical,gibbons2007geometry} for finding such an expansion either use Riemann normal coordinates or exploit very special metrics to find the coefficients of the expansion. The method in this article uses Synge's world function without choosing any particular coordinate system or metric. Finally, I wrap everything up in a proof of \eqref{b-r} in section \ref{curvedcur}. A discussion follows afterward.\\

\section{Causal set d'Alembertian in flat spacetime: an integral equation}
\label{flatflt}
Let $\phi$ be a scalar field with compact support on the 2d flat spacetime, $\mathbb{M}_2$. I want to compute the average of $B\phi(z)$ over all Poisson sprinklings on $\mathbb{M}_2$ with $z$ being added by hand. For a given sprinkling $\mathcal{C}$ on $\mathbb{M}_2$, one can write
$$\sum_{y:|[y,z]|=n}\phi(y)=\int d^2 y\ \chi_{\mathcal{C}}(y)N_n(y,z)\phi(y)$$
where 
$$\chi_{\mathcal{C}}(y)=\sum_{y'\in \mathcal{C}}\delta^2(y-y'), \ \ \ N_n(y,z)=\begin{cases}1,& |[y,z]|=n\\
0,& \text{else}
\end{cases}$$
Taking the expected value, one finds
\begin{equation}
    \mathbb{E}\left[N_n(y,z)\right]={\mathbb{P}}\left(N_n(y,z)=1\right)={\mathbb{P}}\left(|[y,z]|=n\right)=\frac{1}{(n-2)!}e^{-\rho V_{[y,z]}}(\rho V_{[y,z]})^{(n-2)}
\end{equation}
$V_{[y,z]}$ is the volume of the continuum causal diamond $[y,z]=J^-(z)\cap J^+(y)$, and $J^{-}(z)$ is the causal past of $z$.  The effect of $\chi_\mathcal{C}$, being the density of sprinkling, is to introduce $\rho$ when taking the expectation value.
Therefore,
\begin{equation}
    \mathbb{E}\left[\sum_{y:|[y,z]|=2}\phi(y)\right]=\rho\int_{y\preceq z }d^2 y \ e^{-\rho V_{[y,z]}}\phi(y)=\rho\int_{J^{-}(z) }dudv  \ e^{-\rho uv }\phi(u,v)\equiv I_0[\phi].
\end{equation}
For simplicity, I have defined the integral to be $I_0[\phi]$. In the second equality, I have used light-cone coordinates $(u,v)$ in the past of $z$, with $z$ being the origin. 
\begin{align}
    u &= -\frac{1}{\sqrt{2}}(t-x) \\
    v &= -\frac{1}{\sqrt{2}}(t+x).
\end{align}

Similarly, for the second summation in (\ref{Bphi0}):
\begin{equation}
    \mathbb{E}\left[\sum_{y:|[y,z]|=3}\phi(y)\right]=\rho\int_{y\preceq z }d^2 y \ e^{-\rho V_{[y,z]}}\rho V_{[y,z]}\phi(y)=\rho^2\int_{J^{-}(z) }dudv  \ e^{-\rho uv }uv\ \phi(u,v).
\end{equation}
We write $\exp(-\rho uv )uv=-\frac{1}{\rho}\partial_u\left(\exp(-\rho uv)\right)u$, and then integrate by parts:
\begin{flalign}
\mathbb{E}\left[\sum_{y:|[y,z]|=3}\phi(y)\right]&=-\rho\int_{0}^{\infty}\int_{0}^{\infty}dudv\ \partial_{u}\left(e^{-\rho uv }\right)u\phi=\rho\int_0^{\infty}\int_0^{\infty}dudv\ e^{-\rho uv }(\phi+u\partial_u\phi)\nonumber&\\& =I_0[\phi]-\int dudv\ \partial_v\left(e^{-\rho uv }\right)\partial_u\phi=I_0[\phi]+\int_0^{\infty}du\ \partial_u\phi+\int dudv\ e^{-\rho uv }\partial_u\partial_v \phi \nonumber&\\& = I_0[\phi]-\phi(z)+I_1[\phi].
\end{flalign}
Again for simplicity, I have defined
\begin{equation}
    I_1[\phi]\equiv \int dudv\ e^{-\rho uv }\partial_u\partial_v \phi.
\end{equation}
Note that I did not need to set the upper limit of $\int du$ to be $\infty$; any large value greater than the size of the support of $\phi$ works as well.\\
Finally for the third summation in (\ref{Bphi0}), we have
\begin{flalign}
    \mathbb{E}\left[\sum_{y:|[y,z]|=4}\phi(y)\right]&=\frac{1}{2}\rho^3\int_{J^{-}(z) }dudv  \ e^{-\rho uv }(uv)^2\phi(u,v)=-\frac{1}{2}\rho^2\int_{0}^{\infty}\int_{0}^{\infty}dudv\ \partial_{u}\left(e^{-\rho uv }\right)u^2v\phi\nonumber&\\& =\rho^2\int dudv\ e^{-\rho uv }(uv\phi+\frac{1}{2}u^2v\partial_u\phi)&\\&=I_0+I_1-\phi(z)-\frac{\rho}{2}\int dudv\  \partial_{v}\left(e^{-\rho uv }\right)uv\partial_u\phi\nonumber&\\& =I_0+I_1-\phi(z)+\frac{\rho}{2}\int dudv\ e^{-\rho uv }(u\partial_u\phi+uv\partial_u\partial_v\phi).
\end{flalign}
One can go on with this business of getting rid of $u$ and $v$ inside the integral, integrating by parts, and introducing boundary terms. After a few steps, the result is
\begin{equation}
     \mathbb{E}\left[\sum_{y:|[y,z]|=4}\phi(y)\right]=I_0[\phi]+2I_1[\phi]+\frac{1}{2}I_2[\phi]-\frac{3}{2}\phi(z)-\frac{1}{2\rho}\partial_u\partial_v\phi(z),
\end{equation}
where I define 
\begin{equation}
    I_2[\phi]\equiv \frac{1}{\rho}\int dudv\ e^{-\rho uv }\partial^2_u\partial^2_v \phi.
\end{equation}
Now observe that $\partial_u\partial_v\phi(z)=-\frac{1}{2}\Box\phi(z)$. So our desired d'Alembertian has appeared in the third layer of the linear combinations of $\phi$. The other terms, $I_0,\ I_1,\ I_2$ are integrals that can no longer be simplified. Hence, we have
\begin{equation}
    \mathbb{E}[B\phi](z)=4\rho\left((-A+C-\frac{3}{2}D)\phi(z)+(B-C+D)I_0[\phi]+(-C+2D)I_1[\phi]+\frac{D}{2}I_2[\phi]+\frac{D}{4\rho}\Box\phi(z)\right).
\end{equation}
By a suitable choice of the coefficients $A,\ B, \ C,\ D$, we want to set the coefficient of $\Box\phi$ to become one, along with the cancellation of $I_0$, $I_1$, and $\phi(z)$. We do not have the freedom to cancel $I_2$. Therefore, we select 
\begin{equation}
    A=\frac{1}{2},\ \ B=1, \ \ C=2,\ \ D=1.
\end{equation}
Also, keep in mind that $\rho=1/l^2$. So the causal set box operator is 
\begin{equation}
\label{Bphi}
    B\phi(z):=\frac{4}{l^2}\left(-\frac{1}{2}\phi(z)+\sum_{y:|[y,z]|=2}\phi(y)-2\sum_{y:|[y,z]|=3}\phi(y)+\sum_{y:|[y,z]|=4}\phi(y)\right),
\end{equation}
and the above calculations have shown that its expectation value gives
\begin{equation}
\label{box2}
    \mathbb{E}[B\phi](z)=\Box \phi(z)+\frac{1}{2}\int_{J^{-}(z)}d^2y\ e^{-\rho V_{[y,z]} }\Box\Box\phi(y).
\end{equation}
This is a novel exact result in flat spacetime. We can see that the expectation value of the causal set box operator gives the continuum box operator as $\rho\rightarrow\infty$ because the factor $\exp(-\rho V)$ becomes zero in this limit everywhere except \textit{on} the past light cone of $z$. $\Box\Box\phi$ remains finite everywhere including on the light cone, and $\phi$ has compact support, so the integral goes to zero in the limit.\footnote{Note that a similar conclusion could not be made if instead we had $\int d^2y\ e^{-\rho V_{[y,z]} }\rho\Box\phi(y)$ on the right-hand side because $\rho e^{-\rho V_{[y,z]}}$ diverges on the light cone in the infinite density limit.} This equation is favorable from another aspect, too. In \cite{sorkin2009does,aslanbeigi2014generalized} Sorkin, Aslanbeigi, and Saravani argue that the zero eigenmodes of $\Box$ and $\mathbb{E}[B]$ are the same, in order to conclude that an evolution defined by $\mathbb{E}[B\phi]=0$ is stable. One side of this result is now obvious from (\ref{box2}): All massless scalar fields propagating freely in flat spacetime also satisfy the “averaged causal set wave equation", $\mathbb{E}[B\phi]=0$. For the other side, we need to invert \eqref{box2} and expand $\Box\phi$ as an integral series expansion in $\mathbb{E}[B\phi]$. Assuming that the resulting series is convergent, the method of successive substitution (that is typically used to invert Volterra integral equations) gives a nested integral equation:
\begin{equation}
\begin{aligned}[b]
\label{nested}
    \Box \phi(z)=&\ \ \mathbb{E}[B\phi](z)-\frac{1}{2}\int_{y\preceq z}d^2y\ e^{-\rho V_{[y,z]} }\Box_y\mathbb{E}[B\phi]\\&
    +\frac{1}{4}\int_{y_2\preceq y_1\preceq z}d^2y_1 d^2y_2\ e^{-\rho V_{[y_1,z]} }\Box_{y_1}\Big( e^{-\rho V_{[y_2,y_1]}}\Box_{y_2}\mathbb{E}[B\phi]\Big)\\ &
    -\frac{1}{8}\int_{y_3\preceq y_2\preceq y_1\preceq z}d^2y_1 d^2y_2d^2y_3\ e^{-\rho V_{[y_1,z]} }\Box_{y_1}\Bigg( e^{-\rho V_{[y_2,y_1]}}\Box_{y_2}\Big(e^{-\rho V_{[y_3,y_2]}}\Box_{y_3}\mathbb{E}[B\phi] \Big)\Bigg)+\dots\\&
    =\mathbb{E}[B\phi](z)+\sum_{n=1}^{\infty}\left(-\frac{1}{2}\right)^n\int_{y_n\preceq\dots\preceq y_1\preceq z}d^2y_n\dots d^2y_1\  O_{y_1z}\dots O_{y_ny_{n-1}}\mathbb{E}[B\phi](y_n)\,,
\end{aligned}
\end{equation}
where I have defined $O_{xy}=\exp\left(-\rho V_{[x,y]}\right)\Box_{x}$. This completes the proof that
\begin{equation}
    \mathbb{E}[B\phi]=0\ \ \iff\ \   \Box\phi=0.
\end{equation}
Therefore, on average, the causal set wave equation has the same solutions as the Minkowski wave equation in 2d, even at finite density.

The other thing to be learned from this section is the method that is used to compute the expected values. It essentially consisted of replacing the factors $u$ and $v$ inside the integral, one by one, by derivatives of the exponential factor $\exp(-\rho uv)$ and then using successive integration by parts to put the derivatives on $\phi$. I will carry out the same procedure in curved spacetime but with $\sigma$ instead of $uv$, where $\sigma$ is Synge's world function.

\section{Synge's world function}
\label{Syngesin}
Here, for the sake of completeness, I present a brief overview of Synge's world function, $\sigma(y,z)$, and its main properties \cite{synge1960relativity,de1963dynamical,mashhoon2017nonlocal,poisson2011motion,vines2015geodesic}. As we shall shortly see, there are identities governing the covariant derivatives of this bi-scalar that include the Riemann tensor of the manifold. On the other hand, with some effort, I will show in the next section that we can relate $\sigma(y,z)$ to the volume of the causal interval between $y$ and $z$, $V_{[y,z]}$. This volume, according to the Poisson distribution, is the natural quantity that shows up when taking the average of combinatorial quantities over all Poisson sprinklings on a given spacetime. Hence, Synge's world function will be the key link between the combinatorics of causal sets and the geometry and curvature of Lorentzian manifolds. 

The definition of $\sigma(y,z)$ in dimension $d$ goes as follows: Take a geodesic between $y$ and $z$ where $y\preceq z$. I shall be working in a small enough region of the manifold so that such a geodesic is unique. I use $\lambda\in [\lambda_0,\lambda_1]$ as the affine parameter along this geodesic ($x(\lambda_0)=y$), and $u^\mu=dx^\mu/d\lambda$ as the velocity vector, so that $u^\mu u_\mu=-1$. Then:
\begin{equation}
    \sigma (y,z)\equiv-\frac{1}{2}(\lambda_1-\lambda_0)\int_{\lambda_0}^{\lambda_1}d\lambda\  g_{\mu\nu}\big(x(\lambda)\big)u^\mu u^\nu=\dfrac{1}{2}\left( \lambda _{1}-\lambda _{0}\right) ^{2}.
\end{equation}
So it is defined as the half of the geodesic distance squared between two points. The definition also applies to spacelike separations, but it is not of interest in this article. Note that unlike the usual convention and for later convenience, I defined $\sigma$ such that it is positive for timelike separated points. It also follows from the definition that $\sigma$ is zero for null separated points.

Using the definition of $\sigma$, we can apply a small variation $\delta z$ to one of the endpoints of the geodesic. This changes the geodesic and its tangent vectors infinitesimally as
\begin{equation}
\begin{aligned}[b]
\delta \sigma & =-\dfrac{1}{2}\Delta \lambda \int d\lambda\ \  g_{\mu \nu ,\rho }\delta x^{\rho }u^{\mu }u^{\nu }-\Delta \lambda \int d\lambda\ g_{\mu \nu }u^{\mu }\dfrac{d}{d\lambda }\delta x^{\nu }\\
 & =-\Delta\lambda \int d\lambda\left( \Gamma _{\mu\rho }^{\nu }u^{\rho }u_{\nu }\delta x^{\mu}-\dfrac{d}{d\lambda }\left( u_{\mu }\right) \delta x^{\mu }\right) -\Delta \lambda \left( g_{\mu \nu }u^{\mu }\delta x^{\nu}\right) | _{\lambda _{0}}^{\lambda_1}\\
& = -\Delta\lambda u_{\mu }\delta z^{\mu }.
\end{aligned}
\end{equation}
Note that $\Delta\lambda=\lambda_2-\lambda_1$. Now this gives us the derivative of Synge's world function with respect to one of its arguments:
\begin{equation}
 \label{prin}
\sigma _{,\mu }=-\Delta \lambda u_{\mu }\Rightarrow \sigma _{,\mu }\sigma ^{,\mu }=-2\sigma .
\end{equation}
As a note on convention, I use $\sigma _{;\mu_1\dots\mu_j }=\sigma _{\mu_1\dots\mu_j }$ for derivatives with respect to $z$ (i.e. with respect to the upper point), and $\sigma _{;\mu_1'\dots\mu_j' }=\sigma _{\mu_1'\dots\mu_j'}$ for derivatives with respect to $y$. The above equation is saying that $\sigma _{\mu }$ is a vector at $z$, tangent to the geodesic from $y$ to $z$, and pointing toward $y$, with the norm equal to the geodesic distance from $y$ to $z$. Also, it is obviously a scalar with respect to $y$.

For our purpose, we need information about the higher derivatives of $\sigma$ as well. More specifically, we are interested in the expansion of $\Box \sigma=\sigma^{;\mu}_{\hspace{2mm}\mu}$ in terms of geometric quantities of the manifold at $z$. As we shall see, one can derive 
\begin{equation}
    \begin{aligned}\square_z \sigma =-2-\dfrac{1}{3}R\sigma-\dfrac{1}{12}R_{;\mu }\sigma ^{\mu }\sigma -\dfrac{1}{60}R_{;\mu \nu }\sigma^{\mu }\sigma^{\nu}\sigma 
+\dfrac{1}{45}R^{2}\sigma ^{2}-\dfrac{1}{360}R_{;\mu \nu \lambda }\sigma ^{\mu }\sigma ^{\nu }\sigma ^{\lambda }\sigma +\dfrac{1}{60}RR_{;\mu }\sigma ^{\mu }\sigma ^{2}+\dots\end{aligned}
\end{equation}
in 2d. In this equation, all curvature terms are evaluated at $z$, and all world functions have the argument $(y,z)$. If we integrate this equation over a causal interval, the constant term on the right-hand side gives us a term proportional to the volume of the interval. That is why we are interested in such equations. Also, note that in such expansions, we assume that there exists some scale $L_G$ such that $R\sim 1/L_G^2$, and that $L_G$ is also the scale of variations of $R$ so that there exists a coordinate system in which $R_{;\mu}\sim 1/L_G^3$. The small parameter of the expansions is $L/L_G$, where $L$ is the scale of geodesic distance between $y$ and $z$.\\
So we are looking for an equation like
\begin{equation}
 \label{hi}
    \begin{aligned}\sigma _{\mu \nu }=A_{\mu u}+B_{\mu\nu \alpha \beta }\sigma ^{\alpha }\sigma ^{\beta }+C_{\mu \nu \alpha \beta\theta }\sigma ^{\alpha }\sigma ^{\beta }\sigma ^{\theta }
+D_{\mu \nu\alpha\beta\theta \kappa}\sigma ^{\alpha }\sigma ^{\beta }\sigma ^{\theta }\sigma ^{\kappa}+\dots\end{aligned}
\end{equation}
for tensors $A, B, C, D$ at $z$ that are yet to be found. Again note that this is a scalar equation at $y$.
For finding the above unknown tensors, we take the coincidence limit of equation (\ref{hi}) and its covariant derivatives; i.e. we set $y=z$. Quantities that are evaluated at the coincidence limit are put in brackets. For example, the coincidence limit of equation (\ref{prin}) gives $[\sigma_{\mu}]=0$, and this implies for (\ref{hi}) that $[\sigma_{\mu\nu}]=A_{\mu\nu}$. For finding $[\sigma_{\mu\nu}]$, we take a covariant derivative of (\ref{prin}) to find $    \sigma ^{\mu }\sigma _{\mu \nu }=-\sigma _{\nu }$, and another covariant derivative gives
\begin{equation}
\sigma ^{\mu }\sigma _{\mu \nu \alpha }+\sigma _{\hspace{2mm}\alpha }^{\mu }\sigma _{\mu \nu }=-\sigma _{\nu\alpha }.
\end{equation}
The coincidence limit of this equation gives $\left[ \sigma _{\nu \mu }\right] g^{\mu \beta}\left[ \sigma _{\beta \alpha }\right] =-\left[ \sigma _{\nu \alpha }\right] $, and assuming that $[\sigma_{\mu\nu}]$ is an invertible matrix, we find
\begin{equation}
    [\sigma_{\mu\nu}]=-g_{\mu\nu}
\end{equation}
This gives us the first term in the expansion (\ref{hi}). The logic for finding the rest of the coefficients is the same: Repeatedly take covariant derivatives from (\ref{prin}), use the Ricci identity for the commutation of the covariant derivatives, and take the coincidence limit. For the first next steps, one finds
\begin{equation}
\left[ \sigma _{\mu \alpha \beta }\right] =0;\hspace{5mm}\left[ \sigma _{\mu \nu \alpha \beta }\right] =\dfrac{1}{3}\left( R_{\beta\mu \alpha
\nu }+R_{\alpha \mu \beta \nu}\right)
\end{equation}
\begin{equation}
\left[ \sigma _{\mu\nu \alpha \beta \theta}\right] =\dfrac{1}{4}( R_{\alpha \mu \theta\nu ;\beta }+R_{\alpha \mu \beta
\nu ;\theta}+R_{\theta\mu \alpha\nu ;\beta }
+R_{\beta \mu\alpha\nu;\theta}+R_{\beta\mu\theta\nu ;\alpha }+R_{\theta\mu \beta \nu ;\alpha }) .
\end{equation}
The expression for $\sigma$ with 6 indices at coincidence limit is much more complicated and tedious to derive, see \cite{decanini2006off}. Next, take repeated derivatives from (\ref{hi}), then take the coincidence limit, and use the above expressions to determine the coefficients. The result is \cite{decanini2006off}:
\begin{equation}
\begin{aligned}[b]
\label{sigmamn}
   \sigma _{\mu \nu }= &-g_{\mu \nu }+\dfrac{1}{3}R_{\mu \alpha \nu\beta }\sigma ^{\alpha }\sigma ^{\beta }+\dfrac{1}{12}R_{\mu \alpha \nu \beta ;\gamma }\sigma ^{\alpha }\sigma ^{\beta }\sigma ^{\gamma }\\
& +(\dfrac{1}{60}R_{\mu \alpha \nu\beta ;\gamma \delta}+\dfrac{1}{45}R_{\lambda \alpha \mu \beta }R^{\lambda }_{\hspace{1mm}\gamma\nu \delta})\sigma^{\alpha} \sigma^{\beta }\sigma ^{\gamma }\sigma ^{\delta }\\& +\left( \dfrac{1}{360}R_{\mu \alpha \nu \beta ;\gamma \delta \theta }+\dfrac{1}{120}R_{\lambda \alpha \mu \beta }R_{\gamma\nu\delta ;\theta }^{\lambda }+\dfrac{1}{120}R_{\lambda \alpha \mu \beta ;\theta}R^{\lambda }_{\gamma\nu\delta} \right) 
\sigma ^{\alpha }\sigma ^{\beta }\sigma ^{\gamma }\sigma ^{\delta }\sigma ^{\theta }+\dots   
\end{aligned}
\end{equation}
So far, the expressions hold for any dimension. In 2d, the Riemann tensor has a simple form:
\begin{equation}
    R_{\mu\nu\alpha\beta}=\frac{1}{2}(g_{\mu\alpha}g_{\nu\beta}-g_{\mu\beta}g_{\nu\alpha})R.
\end{equation}
This simplifies (\ref{sigmamn}) to
\begin{equation}
\begin{aligned}[b]
   \sigma _{\mu \nu }= &-g_{\mu \nu }-\dfrac{1}{3}g_{\mu \nu}R\sigma-\dfrac{1}{6}R\sigma_{\mu}\sigma_{\nu}-\dfrac{1}{12}R_{;\alpha}g_{\mu\nu}\sigma\sigma^{\alpha}-\dfrac{1}{24}R_{;\alpha}\sigma_\mu\sigma_\nu\sigma^{\alpha }+\dots  
\end{aligned}
\end{equation}

I will be using equation (\ref{sigmamn}) in the case that $\mu$ and $\nu$ are contracted and $d=2$. Under these circumstances, (\ref{sigmamn}) will give 
\begin{equation}
\label{bx}
    \Box_z\sigma=-2-\frac{1}{3}R\sigma-\frac{1}{12}R_{;\mu}\sigma^{\mu}\sigma-\frac{1}{60}R_{;\mu\nu}\sigma^{\mu}\sigma^{\nu}\sigma+\frac{1}{45}R^2\sigma^2-\frac{1}{360}R_{;\mu\nu\alpha}\sigma^{\mu}\sigma^{\nu}\sigma^{\alpha}\sigma+\frac{1}{60}RR_{;\mu}\sigma^{\mu}\sigma^2+\dots
\end{equation}

\section{Expansion of volume of causal intervals}
\label{volvo}
In $\mathbb{M}_2$, the volume of the causal interval between a point $(u,v)$ in light-cone coordinates and the origin is $uv$, which is equal to $\sigma$; therefore, $V[y,z]=\sigma(y,z)$ in flat spacetime.\footnote{This is why I defined $\sigma$ to be positive for timelike separated points; otherwise, a minus sign would have appeared here.} In a 2d curved spacetime, there are corrections of higher powers of $\sigma$ to this equation, with the coefficients being the curvature components of the manifold. As an example of how the expansion formulae for Synge's world function can come in handy, let's see how we can expand $V[y,z]$ in terms of the Riemann curvature at $z$ and $\sigma$. Similar expansions have been previously found in \cite{gibbons2007geometry} in terms of the curvature components at the “center" of the interval. However, I will carry out the expansions about the endpoint of the interval, $z$. At the first order in curvature where one can ignore derivatives and higher powers of the curvature, the two expansions are the same. However, if one would like to think about the continuum limit of the BDG action for causal sets and its higher-order corrections to the Einstein-Hilbert action, then the origin point for the expansion should be one of the endpoints of the interval. The reason is simple: the expectation value of the BDG Lagrangian is an integral $\int d^dy$ over the causal past of the point $z$. Therefore, if what we need is a curvature expansion of such integrals, we would like $V[y,z]$ to be expanded around $z$. The following approach paves the way for such calculations. In addition, the approach used in \cite{gibbons2007geometry} is to write down all possible terms that can appear in the expansion based on dimensional analysis, and find their coefficients by using special 2d metrics for which the volume of a causal interval can be calculated. Here, however, I will not be using any particular coordinate system or metric. This makes it easier to extend the method to find the higher-order terms in the expansion. 

\begin{figure}[H]
\centering
     \includegraphics[width=.8\textwidth]{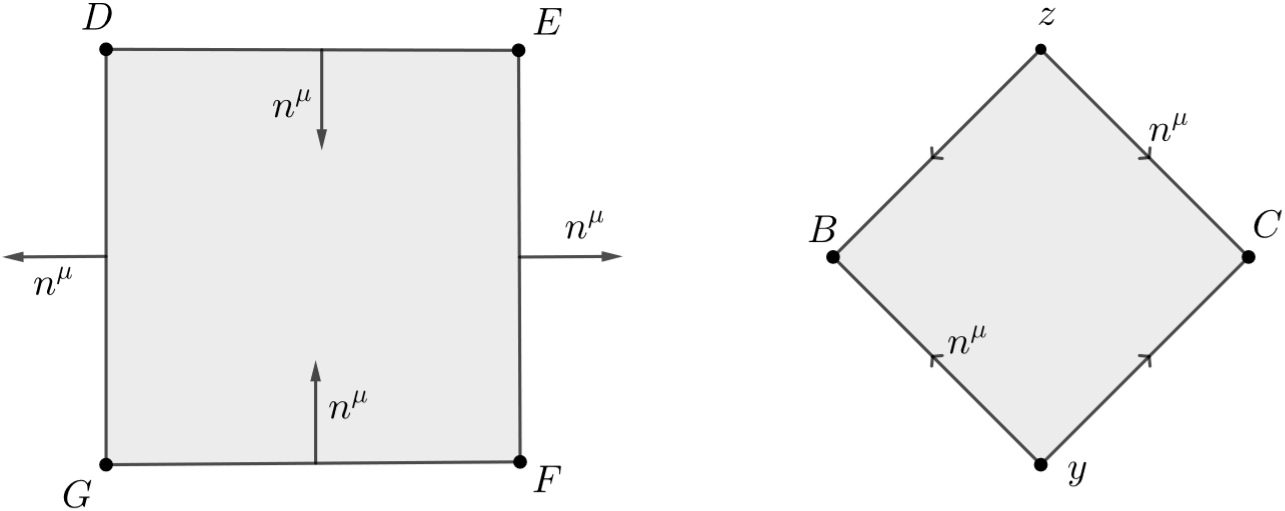}
    \caption{In a Lorentzian manifold, the normal vectors must be outward-pointing if spacelike and inward-pointing if timelike when applying the Stokes theorem, as shown in the left figure. A continuous deformation of the left figure to the right figure tells us how to select the direction of normal vectors when the boundary is null: they should point towards the “middle" of the region, i.e. towards $B$ and $C$. In particular, although the boundary $ByCz$ itself does not have a topological boundary, total derivative integrals on $ByCz$ (like \eqref{td}) do not vanish. They give contributions at the vertices $B$, $y$, $C$, and $z$.}
\label{stoke}
\end{figure}

In the beginning, we need a simple lemma to relate the bulk integrals on causal diamonds to scalar and bi-scalar values at the endpoints.\\
\begin{lemma}
 Let $\psi$ be a scalar function on the 2d Lorentzian manifold $\mathcal{M}$, and let $y,z\in\mathcal{M}$, $y\prec z$ be two points such that $\sigma(y,x)$ is well-defined for all $x\in[y,z]$. Then
 \begin{equation}
 \label{lem}
     \sigma(y,z)\psi(z)=\frac{-1}{2}\int_{[y,z]} d^2x\sqrt{-g}\ \Box_{x}\left(\sigma(y,x)\psi(x)\right).
 \end{equation}
\end{lemma}
The proof first uses Stokes theorem to write the right-hand side as a boundary integral
\begin{equation}
\label{td}
    \frac{-1}{2}\int_{\partial[y,z]} d\Sigma^\mu\ \nabla_\mu\left(\sigma\psi \right),
\end{equation}
where it is understood that the argument for $\sigma$ inside the integral is again $(y,x)$, where $x$ is the boundary point being integrated. Now in a 2-dimensional setting, the boundary integral is equal to (check \cite{wald2010general} for a correct statement of the Stokes theorem in Lorentzian manifolds and see \cite{lehner2016gravitational} for boundary integral techniques)
\begin{equation}
    \frac{-1}{2}\int_{\partial[y,z]} d\lambda\ \frac{\partial}{\partial\lambda}\left(\sigma\psi \right),
\end{equation}
where lambda is the null parametrization of the boundary. $\partial[y,z]$ consists of four null curves, each one giving a contribution from each of their endpoints to the integral, see Figure~\ref{stoke}. Three vertices, including $y$, are at zero geodesic distance to $y$ and so have zero contribution. $z$ lies on the intersection of two null lines and that just gives us (\ref{lem}). This concludes the proof of the lemma.

Now apply the lemma for $\psi(x)=1$. Combining with equation (\ref{bx}), we find
\begin{equation}
\begin{aligned}[b]
\label{Vex}
    \sigma(y,z)&=\frac{-1}{2}\int_{[y,z]} d^2x\sqrt{-g}\ \Box_x\sigma(y,x) \\
    &=\int dV_x+\frac{1}{6}\int dV_x\ R\sigma+\frac{1}{24}\int dV_x\ R_{;\mu}\sigma^\mu\sigma+\frac{1}{120}\int dV_x\ R_{;\mu\nu}\sigma^{\mu}\sigma^{\nu}\sigma\\& -\frac{1}{90}\int dV_x\ R^2\sigma^2+\frac{1}{720}\int dV_x\ R_{;\mu\nu\alpha}\sigma^{\mu}\sigma^{\nu}\sigma^{\alpha}\sigma-\frac{1}{120}\int dV_x \ RR_{;\mu}\sigma^{\mu}\sigma^{2}+\dots
\end{aligned}
\end{equation}
Where $dV_x=d^2x\sqrt{-g(x)}$, and it is understood that the argument for $\sigma$ inside the integrals is $(y,x)$ and that the region of all integrals is $[y,z]$.

Now here is the key idea: The first term on the right-hand side of the above equation is $V[y,z]$, which is our desired non-local quantity that we want to expand around $z$. To eliminate the other integrals, we apply Lemma 1 for all other possible dimensionless combinations of $\sigma(y,z)$ and $R(z)$ that appear inside the integrals; i.e. let's apply the lemma for $\psi(x)= R(x)\sigma(y,x),$ $R_{;\mu}(x)\sigma^{\mu}(y,x)\sigma(y,x),$ $R^2(x)\sigma^2(y,x),$ etc. 
Using the equations presented in section~\ref{Syngesin}, we find
\begin{flalign}
    R(z)\sigma^2(y,z)
    &= \ 4\int dV_x\ R\sigma-2\int dV_x\ R_{;\mu}\sigma^\mu\sigma\nonumber&\\&+\frac{1}{3}\int dV_x\ R^2\sigma^2-\frac{1}{2}\int dV_x\ \Box_x R\ \sigma^2+\frac{1}{12}\int dV_x\ RR_{;\mu}\sigma^{\mu}\sigma^{2}+\dots&\\
    R_{;\mu}(z)\sigma^{\mu}(y,z)\sigma^{2}(y,z)&=\ 6\int dV_x\ R_{;\mu}\sigma^\mu\sigma+\int dV_x\ \Box_x R \ \sigma^2\nonumber&\\ & -2\int dV_x\ R_{;\mu\nu}\sigma^{\mu}\sigma^{\nu}\sigma-\frac{1}{2}\int dV_x \ \nabla_{\mu}\left(\Box_x R\right)\sigma^{\mu}\sigma^{2}+\dots&\\
    R^2(z)\sigma^{3}(y,z)&=-6\int dV_x\ RR_{;\mu}\sigma^{\mu}\sigma^{2}+9\int dV_x\ R^2\sigma^2+\dots&\\
    \Box_z R(z)\ \sigma^{3}(y,z)&= \ 9\int dV_x\ \Box_x R \ \sigma^2-3\int dV_x \ \nabla_{\mu}\left(\Box_x R\right)\sigma^{\mu}\sigma^{2}+\dots&\\
   R_{;\mu\nu}(z)\sigma^\mu\sigma^\nu\sigma^2(y,z)&= \ 8\int dV_x\ R_{;\mu\nu}\sigma^{\mu}\sigma^{\nu}\sigma-\int dV_x\ \Box_x R \ \sigma^2+\int dV_x\ RR_{;\mu}\sigma^{\mu}\sigma^{2}\nonumber&\\&+2\int dV_x \ \nabla_{\mu}\left(\Box_x R\right)\sigma^{\mu}\sigma^{2}-2\int dV_x \ R_{;\mu\nu\alpha}\sigma^\mu\sigma^\nu\sigma^\alpha\sigma+\dots&\\
   R(z)R_{;\mu}(z)\sigma^{\mu}\sigma^3(y,z)&=\ 12\int dV_{x}\ RR_{;\mu}\sigma^{\mu}\sigma^2+\dots&\\
   \nabla_{\mu}\left(\Box_z R(z)\right)\sigma^{\mu}\sigma^3(y,z)&=\ 12\int dV_{x}\  \nabla_{\mu}\left(\Box_x R\right)\sigma^{\mu}\sigma^2+\dots&\\
   \label{last}
   R_{;\mu\nu\alpha}(z)\sigma^{\mu}\sigma^{\nu}\sigma^{\alpha}\sigma^2(y,z)&=\ 10\int dV_x\ R_{;\mu\nu\alpha}\sigma^{\mu}\sigma^{\nu}\sigma^{\alpha}\sigma-\int dV_x\ RR_{;\mu}\sigma^{\mu}\sigma^{2}\nonumber&\\&-3\int dV_x \ \nabla_{\mu}\left(\Box_x R\right)\sigma^{\mu}\sigma^{2}+\dots
\end{flalign}
In deriving these equations, one has to use the following expansions 
\begin{align}\Box_x\sigma^\mu(y,x)&=\frac{1}{6}R(x)\sigma^\mu(y,x)+\mathcal{O}\left(L^2/L_G^3\right);\\
\Box_x\left(\sigma^2(y,x)\right)&=-8\sigma(y,x)-\frac{2}{3}R(x)\sigma^2(y,x)-\frac{1}{6}R_{;\mu}(x)\sigma^\mu\sigma^2(y,x)+\mathcal{O}\left(L^6/L_G^4\right);\\
\Box_x\left(\sigma^3(y,x)\right)&=-18\sigma^2(y,x)+\mathcal{O}\left(L^6/L_G^2\right).
\end{align}
A linear combination of equations (\ref{Vex}-\ref{last}) with coefficients $a$, $b$, etc. gives 
\begin{equation}
\begin{aligned}
    \sigma+a\ R\sigma^2&+b\ R_{;\mu}\sigma^{\mu}\sigma^{2}+c\ R^2\sigma^{3}+d\ \Box_z R\sigma^{3}+e\ R_{;\mu\nu}\sigma^{\mu}\sigma^{\nu}\sigma^{2}\\&+f\ RR_{;\mu}\sigma^{\mu}\sigma^{3}+g\ \nabla_\mu\Box_zR\sigma^{\mu}\sigma^{3}+h\ R_{;\mu\nu\alpha}\sigma^{\mu}\sigma^{\nu}\sigma^{\alpha}\sigma^{2}\\=\ V[y,z]&+\int dV_x\ R\sigma\left(\frac{1}{6}+4a\right)+\int dV_x\ R_{;\mu}\sigma^\mu\sigma\left(\frac{1}{24}-2a+6b\right)\\&+\int dV_x\ R_{;\mu\nu}\sigma^{\mu}\sigma^{\nu}\sigma\left(\frac{1}{120}-2b+8e\right)+\int dV_x\ R^2\sigma^2\left(-\frac{1}{90}+\frac{a}{3}+9c\right)\\&+\int dV_x\ \Box_x R\ \sigma^2\left(-\frac{a}{2}+b+9d-e\right)+\int dV_x\ R_{;\mu\nu\alpha}\sigma^{\mu}\sigma^{\nu}\sigma^{\alpha}\sigma\left(\frac{1}{720}-2e+10h\right)\\&+\int dV_x \ RR_{;\mu}\sigma^{\mu}\sigma^{2}\left(-\frac{1}{120}+\frac{a}{12}-6c+e+12f-h\right)\\&+\int dV_x \ \nabla_{\mu}\left(\Box_x R\right)\sigma^{\mu}\sigma^{2}\left(-\frac{b}{2}-3d+2e+12g-3h\right),
\end{aligned}
\end{equation}
where on the left-hand side, the argument of the world functions is $(y,z)$ and curvature terms are at $z$, while inside the integrals on the right-hand side, the world functions and curvature terms are evaluated at $(y,x)$ and $x$, respectively. Now we can find the coefficients such that only the term $V[y,z]$ survives on the right-hand side. The result is 
\begin{flalign}
\label{volexp}
    V[y,z]&=\sigma(y,z)-\frac{1}{24}R\sigma^2-\frac{1}{48}R_{;\mu}\sigma^\mu\sigma^2+\frac{1}{360}R^2\sigma^3-\frac{1}{160}R_{;\mu\nu}\sigma^\mu\sigma^\nu\sigma^2-\frac{1}{1440}(\Box R)\sigma^3\nonumber&\\
    &-\frac{1}{720}R_{;\mu\nu\alpha}\sigma^\mu\sigma^\nu\sigma^\alpha\sigma^2-\frac{1}{2880}\nabla_{\mu}\left(\Box R\right)\sigma^\mu\sigma^3+\frac{1}{360} RR_{;\mu}\sigma^\mu\sigma^3+\mathcal{O}\left(L^6/L_G^6\right).
\end{flalign}
This equation should be compared with equation (65) in \cite{gibbons2007geometry}. In their expansion, $R$ is evaluated at the center of the interval. Replacing their $R(\text{center})$ with $R(z)+R_{;\mu}(z)\sigma^\mu(y,z)/2+R_{;\mu\nu}(z)\sigma^\mu\sigma^\nu(y,z)/8$ (i.e. with the curvature at the endpoint of the interval) correctly recovers the first line of our equation~\eqref{volexp}.

\section{Application: a covariant approach to causal set action}
\label{curvedcur}
The causal set action was discovered in a series of papers \cite{benincasa2010scalar,belenchia2016continuum,benincasa2013action,dowker2013causal}. It is conjectured that the combinatorial quantity defined by \begin{equation}
\label{actionact}
    \frac{S_{BDG}}{\hbar}=2(N-2N_1+4N_2-2N_3),
\end{equation}is such that its expected value over all Poisson sprinklings of a given 2d spacetime is equal to the Einstein-Hilbert action plus a “corner" boundary term in the large-density limit \cite{dowker2021boundary,machet2020continuum}. More specifically, what is to be proved is that up to the first order in curvature,
\begin{equation}
\label{tobeproved}
    \lim_{\rho\rightarrow\infty}\mathop{\mathbb{E}}[\mathcal{L}](z)=\frac{R(z)}{2}.
\end{equation}
The approach used in \cite{belenchia2016continuum} involves first using the Poisson distribution to write
\begin{equation}
\label{expect}
    \mathop{\mathbb{E}}[\mathcal{L}](z)=2\rho \left(1-2\int_{[y,z]}\rho\ dV_x\ e^{-\rho V_{[x,z]}}+4\int_{[y,z]}\rho^2dV_x\ e^{-\rho V_{[x,z]}}V_{[x,z]}-\int_{[y,z]}\rho^3\ dV_x\ e^{-\rho V_{[x,z]}}V_{[x,z]}^2\right).
\end{equation}
Here it is assumed that the whole region of interest is in the future of some point $y$. The integration region is partitioned into three separate regions, two of which they show give very small contributions. Next, a Riemann normal coordinate system around $z$ is used for the remaining integral. The geometric quantities in this region are expanded in normal coordinates to prove (\ref{tobeproved}). Here I want to show that this third integral over the region close to $z$ can be done without any reference to any coordinate system, by merely using the properties of Synge's world function. So, consider \eqref{expect} where $[y,z]$ is small enough that $\sigma(x,z)$ exists for all $x\in[y,z]$.
Since (\ref{tobeproved}) is a first-order result in curvature, what we need for this particular application is only the $\mathcal{O}\left(L^2/L_G^2\right)$ part of the expansion of $V_{[x,z]}$. First, we need another lemma similar to Lemma 1, but this time including the Poisson factor $\exp(-\rho V_{[x,z]})$.
\begin{lemma}
 Let $\psi$ be a scalar function on the 2d Lorentzian manifold $M$, and let $y,z\in\mathcal{M}$, $y\prec z$. Then
 \begin{flalign}
 \label{lem2}
 \rho^2\int_{[y,z]} dV_x\ e^{-\rho \sigma}\sigma\psi&= -\frac{1}{2} \int_{\partial[y,z]} d \Sigma_{\mu}\ \nabla^{\mu}\left(e^{-\rho \sigma} \psi\right)+\int_{\partial[y,z]}  d\Sigma_{\mu}\ e^{-\rho \sigma} \nabla^{\mu} \psi+\rho \int_{[y,z]} dV_x\ e^{-\rho \sigma} \psi\nonumber&\\&-\frac{1}{2} \int_{[y,z]} dV_x\ e^{-\rho \sigma} \Box_x \psi\nonumber-\frac{R}{12 \rho} \int_{\partial[y,z]} d \Sigma_{\mu}\ \nabla^{\mu}\left(e^{-\rho \sigma} \psi\right)+\frac{R}{6 \rho} \int_{\partial[y,z]} d \Sigma^{\mu}\ e^{-\rho \sigma} \nabla_{\mu} \psi&\\&+\frac{R}{6} \int_{[y,z]} dV_x\ e^{-\rho \sigma} \psi
-\frac{R}{12\rho}  \int_{[y,z]} dV_x\ e^{-\rho \sigma} \square_x \psi +\dots,
 \end{flalign}
 where it is understood that inside the integral, all the bi-scalars $\sigma$ have the argument $(x,z)$ and all scalars $\psi$ are evaluated at $x$, but each $R$ is evaluated at $z$. Derivatives are with respect to $x$. Three dots indicate higher-order curvature terms.
\end{lemma}
To prove this lemma, we replace $\exp(-\rho \sigma)\sigma$ with $-\frac{1}{2}\exp(-\rho \sigma)\sigma_{\mu'}\sigma^{\mu'}$ whenever we see $\sigma$ inside the integral (but not in the exponential). Then write $\exp(-\rho \sigma)\sigma_{\mu'}=-\frac{1}{\rho}\nabla_\mu \left(\exp(-\rho \sigma)\right)$ whenever $\sigma_{\mu'}$ appears in the integral. Then we integrate by parts and keep track of the boundary terms. We use \eqref{bx}, and finally, since we are working only up to the first order terms in curvature, we can replace $R(x)$ by $R(z)$ and bring it out of the integrals. A few steps will lead to \eqref{lem2}.

Let's get to the proof of (\ref{tobeproved}). Take the first integral in (\ref{expect}), and use (\ref{volexp}) to replace $V$ by $\sigma$.
\begin{equation}
    J_0\equiv \rho\int_{[y,z]}dV_x\ e^{-\rho V_{[x,z]}}=\rho\int_{[y,z]}dV_x\ e^{-\rho\sigma_{[x,z]}}(1+\frac{\rho}{24}R\sigma^2)+\mathcal{O}(L^3/L_G^3),
\end{equation}
where $R$ is evaluated at $z$, and so it comes out of the integral.
Do the same for the two other terms
\begin{flalign}
\label{j1}
    J_1&\equiv \rho^2\int_{[y,z]}dV_x\ e^{-\rho V_{[x,z]}}V_{[x,z]}=\rho^2\int_{[y,z]}dV_x\ e^{-\rho\sigma}(1+\frac{\rho}{24}R\sigma^2)(\sigma-\frac{1}{24}R\sigma^2)+\dots\nonumber&\\&= \rho^2\int_{[y,z]}dV_x\ e^{-\rho \sigma}\sigma+\frac{\rho^3}{24}\int_{[y,z]}dV_x\ e^{-\rho \sigma}R\sigma^3-\frac{\rho^2}{24}\int_{[y,z]}dV_x\ e^{-\rho \sigma}R\sigma^2+\dots&\\
    \label{j2}
    J_2&\equiv \frac{\rho^3}{2}\int_{[y,z]}dV_x\ e^{-\rho V_{[x,z]}}V_{[x,z]}^2=\frac{\rho^3}{2}\int_{[y,z]}dV_x\ e^{-\rho\sigma}(1+\frac{\rho}{24}R\sigma^2)(\sigma-\frac{1}{24}R\sigma^2)^2+\dots\nonumber&\\&= \frac{\rho^3}{2}\int_{[y,z]}dV_x\ e^{-\rho \sigma}\sigma^2+\frac{\rho^4}{48}\int_{[y,z]}dV_x\ e^{-\rho \sigma}R\sigma^4-\frac{\rho^3}{24}\int_{[y,z]}dV_x\ e^{-\rho \sigma}R\sigma^3+\dots
\end{flalign}
Integrals that appear in the above expressions can be evaluated using Lemma 2. For example, for the first term on the right-hand side of (\ref{j1}), use Lemma 2 with $\psi=1$ to obtain
\begin{flalign}
I_1\equiv \rho^2\int_{[y,z]} dV_x\ e^{-\rho \sigma}\sigma&= -\frac{1}{2} \int_{\partial[y,z]} d \Sigma_{\mu}\ \nabla^{\mu}\left(e^{-\rho \sigma} \right)+I_0-\frac{R}{12 \rho} \int_{\partial[y,z]} d \Sigma_{\mu}\ \nabla^{\mu}\left(e^{-\rho \sigma} \right)\nonumber&\\&+\frac{R}{6} \int_{[y,z]} dV_x\ e^{-\rho \sigma}=(\beta_0+I_0)(1+\frac{R}{6\rho}),
\end{flalign}
where I have defined, for simplicity,
\begin{equation}
     I_0\equiv\rho\int_{[y,z]} dV_x\ e^{-\rho \sigma},\ \ \ \beta_0\equiv-\frac{1}{2} \int_{\partial[y,z]} d \Sigma_{\mu}\ \nabla^{\mu}\left(e^{-\rho \sigma} \right).
\end{equation}
Also for the first term on the right-hand side of (\ref{j2}), use Lemma 2 with $\psi(x)=\rho\sigma(x,z)/2$ to get
\begin{flalign}
\label{i2}
I_2&\equiv\  \frac{\rho^3}{2}\int_{[y,z]} dV_x\ e^{-\rho \sigma}\sigma^2= -\frac{\rho}{4} \int_{\partial[y,z]} d \Sigma_{\mu}\ \nabla^{\mu}\left(e^{-\rho \sigma}\sigma \right) -\frac{1}{2} \int_{\partial[y,z]} d \Sigma_{\mu}\ \nabla^{\mu}\left(e^{-\rho \sigma} \right)\nonumber&\\&+\frac{\rho^2}{2}\int_{[y,z]} dV_x\ e^{-\rho \sigma}\sigma-\frac{\rho}{4}\int_{[y,z]} dV_x\ e^{-\rho \sigma}(-2-\frac{1}{3}R\sigma)-\frac{R}{24 } \int_{\partial[y,z]} d \Sigma_{\mu}\ \nabla^{\mu}\left(e^{-\rho \sigma}\sigma \right)\nonumber&\\&+\frac{R}{12} \int_{\partial[y,z]} d \Sigma_{\mu}\ e^{-\rho \sigma}\nabla^{\mu}\left(\sigma \right)+\frac{\rho R}{12} \int_{[y,z]} dV_x\ e^{-\rho \sigma}\sigma-\frac{R}{24}\int_{[y,z]} dV_x\ e^{-\rho \sigma}(-2-\frac{1}{3}R\sigma),
\end{flalign}
where I have used (\ref{bx}). Again for simplification let's define
\begin{equation}
    \beta '_0\equiv\frac{\rho}{4} \int_{\partial[y,z]} d \Sigma_{\mu}\ \nabla^{\mu}\left(e^{-\rho \sigma}\sigma \right),
\end{equation}
so that (\ref{i2}) can be written as
\begin{equation}
    I_2=-\beta '_0+\beta_0+\frac{1}{2}(I_1+I_0)+\frac{R}{6\rho}I_1-\frac{R}{6\rho}\beta '_0+\frac{R}{6\rho}\beta_0+\frac{R}{12\rho}I_0.
\end{equation}
The same idea is applied to the last two terms in (\ref{j2}). Using Lemma 2 with $\psi(x)=\frac{\rho}{24}R(x)\sigma^2(x,z)\simeq \frac{\rho}{24}R(z)\sigma^2(x,z)$ we find
\begin{equation}
    RI_3\equiv \frac{\rho^3}{24}R\int_{[y,z]} dV_x\ e^{-\rho \sigma}\sigma^3=-R\beta '_1+R\beta_1+\frac{R}{12\rho}I_2+\frac{R}{6\rho}I_1,
\end{equation}
where I have defined
\begin{equation}
    \beta '_1\equiv\frac{\rho}{48} \int_{\partial[y,z]} d \Sigma_{\mu}\ \nabla^{\mu}\left(e^{-\rho \sigma}\sigma^2 \right),\ \
    \beta_1\equiv\frac{\rho}{24} \int_{\partial[y,z]} d \Sigma_{\mu}\ e^{-\rho \sigma} \nabla^{\mu}_x\left(\sigma^2 \right).
\end{equation}
And finally another use of Lemma 2 with $\psi(x)=\frac{\rho^2}{48}R(x)\sigma^3(x,z)$ gives
\begin{equation}
    RI_4\equiv \frac{\rho^4}{48}R\int_{[y,z]} dV_x\ e^{-\rho \sigma}\sigma^4=-R\beta '_2+R\beta_2+\frac{R}{2}I_3+\frac{3R}{8\rho}I_2,
\end{equation}
with the definitions
\begin{equation}
    \beta '_2\equiv\frac{\rho^2}{96} \int_{\partial[y,z]} d \Sigma_{\mu}\ \nabla^{\mu}\left(e^{-\rho \sigma}\sigma^3 \right),\ \
    \beta_2\equiv\frac{\rho^2}{48} \int_{\partial[y,z]} d \Sigma_{\mu}\ e^{-\rho \sigma} \nabla^{\mu}_x\left(\sigma^3 \right).
\end{equation}
Then we can rewrite the expectation value of the Lagrangian as
\begin{flalign}
\label{EE}
     \mathop{\mathbb{E}}[\mathcal{L}](z)&=\ 2\rho\left(1-2J_0+4J_1-2J_2\right)=2\rho\left(1-2I_0+4I_1-2I_2-\frac{R}{2\rho}I_2+6RI_3-2RI_4\right)\nonumber&\\& =\ 2\rho\left(1+3\beta_0-2\gamma_0+R\left(\frac{7}{12\rho}\beta_0-\frac{7}{6\rho}\gamma_0+5\gamma_1-2\gamma_2\right)\right),
\end{flalign}
in which I used the notation $\gamma_i=\beta_i-\beta '_i$. All of the coefficients in the above expression are boundary integrals.\\
Now, similar to the proof for Lemma 1, one can show that for $i=0,1,2$, we have $\beta '_i\approx 0$, where “$\approx$" means equal up to an exponentially suppressed difference in the small parameter $l/L$. For example, for $i=0$
\begin{equation}
    \beta '_0=\frac{\rho}{4} \int_{\partial[y,z]} d \Sigma_{\mu}\ \nabla^{\mu}\left(e^{-\rho \sigma}\sigma \right)=\frac{\rho}{4} \int_{\partial[y,z]} d \lambda\ \frac{\partial}{\partial \lambda}\left(e^{-\rho \sigma}\sigma \right)=-\frac{\rho}{2}e^{-\rho\sigma(y,z)}\sigma(y,z).
\end{equation}

Now let's get to $\beta_i$:
\begin{flalign}
\gamma_2\approx \beta_2 &=\frac{\rho^2}{48} \int_{\partial[y,z]} d \Sigma_{\mu}\ e^{-\rho \sigma} \nabla^{\mu}_x\left(\sigma^3 \right) =\frac{\rho^2}{16} \int_{\partial[y,z]} d \Sigma_{\mu}\ e^{-\rho \sigma} \sigma^2\sigma^\mu    =-\frac{\rho}{16} \int_{\partial[y,z]} d \Sigma_{\mu}\ \nabla^\mu\left(e^{-\rho \sigma}\right) \sigma^2\nonumber &\\&=-\frac{\rho}{16} \int_{\partial[y,z]} d \Sigma_{\mu}\ \nabla^\mu\left(e^{-\rho \sigma}\sigma^2\right)   +\frac{\rho}{16} \int_{\partial[y,z]} d \Sigma_{\mu}\ e^{-\rho \sigma}\nabla^\mu_x\left(\sigma^2\right)  \approx\ \frac{3}{2}\beta_1&\\
\gamma_1\approx \beta_1 &=\frac{\rho}{24} \int_{\partial[y,z]} d \Sigma_{\mu}\ e^{-\rho \sigma} \nabla^{\mu}_x\left(\sigma^2 \right) =\frac{\rho}{12} \int_{\partial[y,z]} d \Sigma_{\mu}\ e^{-\rho \sigma} \sigma\sigma^\mu    =-\frac{1  }{12} \int_{\partial[y,z]} d \Sigma_{\mu}\ \nabla^\mu\left(e^{-\rho \sigma}\right) \sigma\nonumber &\\&=-\frac{1}{12} \int_{\partial[y,z]} d \Sigma_{\mu}\ \nabla^\mu\left(e^{-\rho \sigma}\sigma\right)   +\frac{1}{12} \int_{\partial[y,z]} d \Sigma_{\mu}\ e^{-\rho \sigma}\nabla^\mu_x\left(\sigma\right)  \approx\ \frac{1    }{6\rho}\beta_0&\\
\gamma_0\approx\beta_0&=-\frac{1}{2} \int_{\partial[y,z]} d \Sigma_{\mu}\ \nabla^{\mu}\left(e^{-\rho \sigma} \right)=-\frac{1}{2} \int_{\partial[y,z]} d \lambda\ \frac{\partial}{\partial\lambda}\left(e^{-\rho \sigma} \right)=-1+e^{-\rho\sigma(y,z)}\approx-1
\end{flalign}
Therefore we find from (\ref{EE})
\begin{equation}
    \mathop{\mathbb{E}}[\mathcal{L}](z)\approx\frac{R}{2}\left(1+\mathcal{O}(L/L_G)\right).
\end{equation}
There are two types of approximation involved. The first type consists of ignoring all higher order curvature corrections by assuming that $R(z)\sigma(y,z)$ is small, or $L\ll L_G$. Remember that $L$ is the absolute value of the geodesic distance between $y$ and $z$. The second one assumes that $\exp(-\rho\sigma(y,z))\ll 1$, or $l\ll L$. In the limit $\rho\rightarrow\infty$, the second approximation becomes exact. Therefore,
\begin{equation}
\label{limlimm}
    \lim_{\rho\rightarrow\infty}\mathop{\mathbb{E}}[\mathcal{L}](z)=\frac{R(z)}{2}\left(1+\mathcal{O}(L/L_G)\right).
\end{equation}
The smallness of $L/L_G$ is akin to the assumption of the validity of Riemann normal coordinates in \cite{belenchia2016continuum}.

\section{Discussion}
I have provided a new toolkit for dealing with the volume of causal intervals and the expected values usually encountered in causal sets in 2d in a totally covariant manner. The actual calculations might still be long, but have the advantage of avoiding particular coordinate systems or metrics. The approach also gave an interesting integral equation connecting $\Box$ and $\mathbb{E}B$ in flat spacetime, hence proving that the two operators have the same zero modes. This shows that an evolution defined by $B\phi=0$ on a causal set is stable and unitary, at least in an average-over-sprinklings sense. Therefore, it is not possible to derive a swerves-like behavior for massless fields \cite{philpott2009energy} (that includes spontaneous energy gain or loss of massless particles in vacuum) from this kind of scalar field dynamics. For massive scalars, however, things are different. If we impose the dynamical equation $B\phi=m^2\phi$ on a scalar field $\phi$ on a causal set, then in the continuum limit, on average, we would have $\mathbb{E}B\phi=m^2\phi$. Equation~\eqref{nested} then tells us
\begin{equation}
\label{finale}
    \Box \phi(z)=m^2\phi(z)+m^2\sum_{n=1}^{\infty}\left(-\frac{1}{2}\right)^n\int_{y_n\preceq\dots\preceq y_1\preceq z}d^2y_n\dots d^2y_1\  O_{y_1z}\dots O_{y_ny_{n-1}}\phi(y_n).
\end{equation}
This is a causal but non-local equation of motion for $\phi$. In eikonal approximation, one would not recover the dispersion relation $k^2=-m^2$ anymore. This equation (and hence the geodesic equation) will receive non-local corrections. Whether the equation~\eqref{finale} can fully recover a swerves-like diffusion from geodesic motion needs further study. If the answer turns out to be affirmative, it could put the phenomenological swerves model on firmer ground.

Another avenue for further investigation is to include higher-order curvature terms. In principle, one can think of extending the BDG action \eqref{actionact} by adding more layers; i.e. adding terms like $N_i$ for $i>3$ and tuning the coefficients such that \eqref{tobeproved} holds true. However, even if we assume that the current form of the Lagrangian is exact and fundamental, its expected value at finite density $\rho$ contains local higher-order curvature terms suppressed by powers of $l/L_G$ (like $R(z)^2/\rho$) plus some possible “bi-local" terms suppressed by an exponential and powers of $L/L_G$ (like $R(z)^2\sigma(y,z)\exp(-\rho\sigma)$) plus some possible non-local integrals (like $\int dV_x\ \exp(-\rho\sigma)R^2(x)$). Given that we have the expression for $V[y,z]$ up to $\mathcal{O}(RR_{;\mu})$, this is only a matter of a lengthy computation. 

One may ask why I have retained $\left(1+\mathcal{O}(L/L_G)\right)$ in \eqref{limlimm} even after taking the limit $\rho\rightarrow\infty$. This also connects to the above discussion on the form of higher-order corrections. As mentioned in \cite{benincasa2010scalar}, if all the higher order corrections are local terms, they vanish in the $\rho\rightarrow\infty$ limit. However, it is very unlikely not to get any non-local higher order curvature term; we have already seen that even in flat spacetime, there is a non-local term in the expected value of the box operator, \eqref{box2}. This non-local correction, nevertheless, vanishes in the $\rho\rightarrow\infty$ limit. Now if the corrections to \eqref{limlimm} are of the form $R(z)^2\sigma(y,z)\exp(-\rho\sigma)$ or $\int dV_x\ \exp(-\rho\sigma)R^2(x)$, they vanish in the $\rho\rightarrow\infty$ limit, but in general, there may exist non-local corrections (merely based on dimensional grounds) that do not vanish in the limit, like $R(z)^2\sigma(y,z)$ or $\int dV_x\ \rho\exp(-\rho\sigma)R_{;\mu}(x)\sigma^{\mu'}(x,z)$. The proof in \cite{belenchia2016continuum} in 4d (based on expansion in Riemann normal coordinates) suggests that such non-local terms should not appear. It would be interesting to see this explicitly in 2d using the methods introduced in this article.

A natural question is whether this work can be extended to higher dimensions such as 4d. The missing piece is the analogue of Lemma 1 in 4d. A conjecture is
 \begin{equation}
     8\pi\sigma(y,z)^2\psi(z)\stackrel{?}{=}\int_{[y,z]} d^4x\sqrt{-g}\ \Box_{x}\Box_{x}\left(\sigma(y,x)^2\psi(x)\right)
 \end{equation}
Using the volume formula for an interval in $\mathbb{M}_4$ \cite{rideout2006evidence}, one can show that the conjecture is true for flat 4d spacetime and constant $\psi$. Whether it holds in curved spacetime or not is yet unknown.
\section*{Acknowledgments}
I greatly thank Fay Dowker for numerous ideas, comments, and corrections during the completion of this work. I also thank Toby Wiseman for an intriguing discussion on the subject. This research is funded by the President’s PhD Scholarships from Imperial College London.

\bibliographystyle{unsrt}
\bibliography{main.bib}

\begin{thebibliography}{10}

\bibitem{bombelli1987space}
Luca Bombelli, Joohan Lee, David Meyer, and Rafael~D Sorkin.
\newblock Space-time as a causal set.
\newblock {\em Physical review letters}, 59(5):521, 1987.

\bibitem{surya2019causal}
Sumati Surya.
\newblock The causal set approach to quantum gravity.
\newblock {\em Living Reviews in Relativity}, 22(1):1--75, 2019.

\bibitem{sorkin2005causal}
Rafael~D Sorkin.
\newblock Causal sets: Discrete gravity.
\newblock In {\em Lectures on quantum gravity}, pages 305--327. Springer, 2005.

\bibitem{surya2011directions}
Sumati Surya.
\newblock Directions in causal set quantum gravity.
\newblock {\em arXiv preprint arXiv:1103.6272}, 2011.

\bibitem{bombelli2009discreteness}
Luca Bombelli, Joe Henson, and Rafael~D Sorkin.
\newblock Discreteness without symmetry breaking: a theorem.
\newblock {\em Modern Physics Letters A}, 24(32):2579--2587, 2009.

\bibitem{dowker2020symmetry}
Fay Dowker and Rafael~D Sorkin.
\newblock Symmetry-breaking and zero-one laws.
\newblock {\em Classical and Quantum Gravity}, 37(15):155007, 2020.

\bibitem{saravani2014causal}
Mehdi Saravani and Siavash Aslanbeigi.
\newblock On the causal set--continuum correspondence.
\newblock {\em Classical and Quantum Gravity}, 31(20):205013, 2014.

\bibitem{hawking1976new}
Stephen~W Hawking, A~Ro King, and PJ~McCarthy.
\newblock A new topology for curved space--time which incorporates the causal,
  differential, and conformal structures.
\newblock {\em Journal of mathematical physics}, 17(2):174--181, 1976.

\bibitem{malament1977class}
David~B Malament.
\newblock The class of continuous timelike curves determines the topology of
  spacetime.
\newblock {\em Journal of mathematical physics}, 18(7):1399--1404, 1977.

\bibitem{brightwell1991structure}
Graham Brightwell and Ruth Gregory.
\newblock Structure of random discrete spacetime.
\newblock {\em Physical review letters}, 66(3):260, 1991.

\bibitem{rideout2009spacelike}
David Rideout and Petros Wallden.
\newblock Spacelike distance from discrete causal order.
\newblock {\em Classical and quantum gravity}, 26(15):155013, 2009.

\bibitem{benincasa2010scalar}
Dionigi~MT Benincasa and Fay Dowker.
\newblock Scalar curvature of a causal set.
\newblock {\em Physical review letters}, 104(18):181301, 2010.

\bibitem{belenchia2016continuum}
Alessio Belenchia, Dionigi~MT Benincasa, and Fay Dowker.
\newblock The continuum limit of a 4-dimensional causal set scalar
  d’alembertian.
\newblock {\em Classical and Quantum Gravity}, 33(24):245018, 2016.

\bibitem{benincasa2013action}
Dionigi Maria~Teofilo Benincasa.
\newblock The action of a casual set.
\newblock 2013.

\bibitem{sorkin2009does}
Rafael~D Sorkin.
\newblock Does locality fail at intermediate length-scales.
\newblock {\em Approaches to quantum gravity: Toward a new understanding of
  space, time and matter}, pages 26--43, 2009.

\bibitem{skpr}
Rafael~D Sorkin.
\newblock Unpublished notes.
\newblock 2009.

\bibitem{myrheim1978statistical}
Jan Myrheim.
\newblock Statistical geometry.
\newblock Technical report, 1978.

\bibitem{gibbons2007geometry}
GW~Gibbons and SN~Solodukhin.
\newblock The geometry of small causal diamonds.
\newblock {\em Physics Letters B}, 649(4):317--324, 2007.

\bibitem{aslanbeigi2014generalized}
Siavash Aslanbeigi, Mehdi Saravani, and Rafael~D Sorkin.
\newblock Generalized causal set d’alembertians.
\newblock {\em Journal of High Energy Physics}, 2014(6):1--25, 2014.

\bibitem{synge1960relativity}
John~Lighton Synge.
\newblock Relativity: the general theory.
\newblock 1960.

\bibitem{de1963dynamical}
Bryce DE~WITT.
\newblock Dynamical theory of groups and fields.
\newblock {\em Relativity, groups and topology}, 787, 1963.

\bibitem{mashhoon2017nonlocal}
Bahram Mashhoon.
\newblock {\em Nonlocal gravity}, volume 167.
\newblock Oxford University Press, 2017.

\bibitem{poisson2011motion}
Eric Poisson, Adam Pound, and Ian Vega.
\newblock The motion of point particles in curved spacetime.
\newblock {\em Living Reviews in Relativity}, 14(1):1--190, 2011.

\bibitem{vines2015geodesic}
Justin Vines.
\newblock Geodesic deviation at higher orders via covariant bitensors.
\newblock {\em General Relativity and Gravitation}, 47(5):1--28, 2015.

\bibitem{decanini2006off}
Yves Decanini and Antoine Folacci.
\newblock Off-diagonal coefficients of the dewitt-schwinger and hadamard
  representations of the feynman propagator.
\newblock {\em Physical Review D}, 73(4):044027, 2006.

\bibitem{wald2010general}
Robert~M Wald.
\newblock {\em General relativity}.
\newblock University of Chicago press, 2010.

\bibitem{lehner2016gravitational}
Luis Lehner, Robert~C Myers, Eric Poisson, and Rafael~D Sorkin.
\newblock Gravitational action with null boundaries.
\newblock {\em Physical Review D}, 94(8):084046, 2016.

\bibitem{dowker2013causal}
Fay Dowker and Lisa Glaser.
\newblock Causal set d'alembertians for various dimensions.
\newblock {\em Classical and Quantum Gravity}, 30(19):195016, 2013.

\bibitem{dowker2021boundary}
Fay Dowker.
\newblock Boundary contributions in the causal set action.
\newblock {\em Classical and Quantum Gravity}, 38(7):075018, 2021.

\bibitem{machet2020continuum}
Ludovico Machet and Jinzhao Wang.
\newblock On the continuum limit of benincasa--dowker--glaser causal set
  action.
\newblock {\em Classical and Quantum Gravity}, 38(1):015010, 2020.

\bibitem{philpott2009energy}
Lydia Philpott, Fay Dowker, and Rafael~D Sorkin.
\newblock Energy-momentum diffusion from spacetime discreteness.
\newblock {\em Physical Review D}, 79(12):124047, 2009.

\bibitem{rideout2006evidence}
David Rideout and Stefan Zohren.
\newblock Evidence for an entropy bound from fundamentally discrete gravity.
\newblock {\em Classical and quantum gravity}, 23(22):6195, 2006.

\end{thebibliography}

\end{document}